\documentclass[]{aastex62}
\usepackage[]{graphicx}  
\usepackage{enumitem}
 \hypersetup{linkcolor=red,citecolor=blue,filecolor=cyan,urlcolor=magenta}

\submitjournal{Astrophys. J.}

\shorttitle{Alfv\'enic switchbacks emerging from funnels}
\shortauthors{Bale et al.}

\begin{document}

\title{A solar source of Alfv\'enic magnetic field switchbacks:\\{\em in situ} remnants of magnetic funnels on supergranulation  scales}
\correspondingauthor{Stuart D. Bale}
\email{bale@berkeley.edu}

\author[0000-0002-1989-3596]{S. D. Bale}
\affil{Physics Department, University of California, Berkeley, CA 94720-7300, USA}
\affil{Space Sciences Laboratory, University of California, Berkeley, CA 94720-7450, USA}

\author[0000-0002-7572-4690]{T. S. Horbury}
\affil{The Blackett Laboratory, Imperial College London, London, SW7 2AZ, UK}
\author[0000-0002-2381-3106]{M. Velli}
\affil{Earth Planetary and Space Sciences, UCLA, CA  90095, USA}

\author[0000-0002-7318-6008]{M. I. Desai}
\affil{Southwest Research Institute, 6220 Culebra Rd, San Antonio, TX 78238, USA}
\affil{3University of Texas at San Antonio, San Antonio, TX 78249, USA}

\author[0000-0001-5258-6128]{J. S. Halekas}
\affil{Department of Physics and Astronomy, University of Iowa, Iowa City, IA 52242, USA}

\author[0000-0001-6077-4145]{M. D. McManus}
\affil{Physics Department, University of California, Berkeley, CA 94720-7300, USA}
\affil{Space Sciences Laboratory, University of California, Berkeley, CA 94720-7450, USA}

\author[0000-0002-4440-7166]{O. Panasenco}
\affil{Advanced Heliophysics, Pasadena, CA 91106, USA}

\author[0000-0002-6145-436X]{S. T. Badman}
\affil{Physics Department, University of California, Berkeley, CA 94720-7300, USA}
\affil{Space Sciences Laboratory, University of California, Berkeley, CA 94720-7450, USA}

\author[0000-0002-4625-3332]{T. A. Bowen}
\affil{Space Sciences Laboratory, University of California, Berkeley, CA 94720-7450, USA}

\author[0000-0003-4177-3328]{B. D. G. Chandran}
\affil{Department of Physics \& Astronomy, University of New Hampshire, Durham, NH 03824, USA}
\affil{Space Science Center, University of New Hampshire, Durham, NH 03824, USA}

\author[0000-0002-9150-1841]{J. F. Drake}
\affil{Department of Physics, University of Maryland, College Park, MD 20742, USA}
\affil{Institute for Physical Science and Technology, University of Maryland, College Park, MD 20742, USA}
\affil{Joint Space Science Institute, University of Maryland, College Park, MD 20742, USA}

\author[0000-0002-7077-930X]{J. C. Kasper}
\affiliation{BWX Technologies, Inc Washington DC 20002}
\affiliation{Climate and Space Sciences and Engineering, University of Michigan, Ann Arbor, MI 48109, USA}

\author[0000-0002-6577-5515]{R. Laker} 
\affiliation{The Blackett Laboratory, Imperial College London, London, SW7 2AZ, UK}

\author[0000-0001-9202-1340]{A. Mallet}
\affil{Space Sciences Laboratory, University of California, Berkeley, CA 94720-7450, USA}

\author[0000-0002-6276-7771]{L. Matteini}
\affil{The Blackett Laboratory, Imperial College London, London, SW7 2AZ, UK}

\author[0000-0002-6924-9408]{T. D. Phan}
\affil{Space Sciences Laboratory, University of California, Berkeley, CA 94720-7450, USA}

\author[0000-0003-2409-3742]{N. E. Raouafi}
\affiliation{The Johns Hopkins Applied Physics Laboratory,  Laurel, MD 20723, USA}

\author[0000-0001-8479-962X]{J. Squire}
\affiliation{Physics Department, University of Otago, Dunedin 9010, New Zealand}

\author[0000-0003-2845-4250]{L. D. Woodham}
\affil{The Blackett Laboratory, Imperial College London, London, SW7 2AZ, UK}

\author[0000-0002-9202-619X]{T. Woolley}
\affil{The Blackett Laboratory, Imperial College London, London, SW7 2AZ, UK}

\begin{abstract}
One of the striking observations from the Parker Solar Probe (PSP) 
spacecraft is the prevalence in the inner heliosphere of large
amplitude, Alfv\'enic magnetic field reversals termed 'switchbacks'. 
These $\delta B_R/B \sim \mathcal{O}(1$) fluctuations occur on a range
of timescales and in {\em patches}
separated by intervals of quiet, radial magnetic field.
We use measurements from PSP to demonstrate that patches of switchbacks are localized 
within the extensions of plasma structures originating at the base of the corona.  These structures are
characterized by an increase in alpha particle abundance, Mach number,
plasma $\beta$ and pressure, and by depletions in the magnetic field
magnitude and electron temperature.  These intervals are
in pressure-balance, implying stationary spatial structure,
and the field depressions are consistent with
overexpanded flux tubes.  The structures are asymmetric in Carrington
longitude with a steeper leading edge and a small
($\sim$1$^\circ$) edge of hotter plasma and enhanced magnetic field
fluctuations.  Some structures contain suprathermal ions to
$\sim$85 keV that we argue are the energetic tail of the solar wind
alpha population. The structures are separated in longitude by
angular scales associated with supergranulation.  This suggests that these switchbacks originate
 near the leading edge of the diverging magnetic field funnels
associated with the network magnetic field - the
primary wind sources. We propose an origin of the magnetic
field switchbacks, hot plasma and suprathermals, alpha particles 
in interchange reconnection events just above the solar transition
region and our measurements represent the extended regions of a
turbulent outflow exhaust. 
\end{abstract}


\section{Introduction} \label{sec:intro}
The NASA Parker Solar Probe (PSP) mission \citep{2016SSRv..204....7F} was executed to make direct {\em in situ} measurements of the source regions of the solar wind and to understand the heating and acceleration mechanisms using those measurements.  A foundational observation from PSP has been the prevalence of large $\delta B_R/B \sim \mathcal{O}(1)$ Alfv\'enic field reversals that had earlier been termed 'switchbacks' (hereafter 'SBs'), which are interspersed among intervals of 'quiet' primarily radial magnetic field \citep{2019Natur.576..237B, 2019Natur.576..228K, 2020ApJS..246...45H,2020ApJS..246...39D}.  While the SBs had been identified previously using spacecraft measurements at 0.3 AU \citep{2018MNRAS.478.1980H}, at 1 AU \citep{2009ApJ...695L.213G, 2011ApJ...737L..35G} and at high solar latitudes \citep{1999GeoRL..26..631B,2014GeoRL..41..259M}, the PSP measurements are notable for the abundance of events and the apparent patterns of SBs and quiet wind.  Several models of SB generation and evolution have been developed recently with sources in velocity shear in the corona \citep{2005ESASP.592..785L,2020ApJ...902...94R,2021ApJ...909...95S}, magnetic reconnection \citep{2020ApJ...894L...4F, 2021A&A...650A...2D,2020ApJ...903....1Z}, impulsive energy injection within magnetic funnels \citep{2021ApJ...911...75M,2021ApJ...914....8M}, and the nonlinear radial evolution of large amplitude Alfv\'enic fluctuations \citep{2006GeoRL..3314101L,2020ApJ...891L...2S,2021arXiv210408321M,2021arXiv210109529S}.  A key question is whether the SBs are generated at the source of the solar wind, and serve as a fundamental diagnostic of the energization mechanism, or if rather they are a product of radial evolution.  

Because of their ubiquity, and apparent nonlinearity, the possibility that switchbacks provide a direct, {\em in situ} diagnostic of solar wind heating or energization is tantalizing.  A variety of heating mechanisms have been proposed:  nanoflares, wave-dissipation, footpoint shearing, magnetic reconnection, and turbulence, to name a few. Most of these theories find some support in remote-sensing or {\em in situ} data and many of them rely in impulsive magnetic activity at the coronal base.  There are many good review papers on the topic, recently  by \cite{Cranmer.ARAA.2019}.

Here we use PSP measurements to demonstrate that an interval of periodically-modulated SBs observed below 25 solar radii ($R_S$) of PSP Encounter 06 is related directly to underlying photospheric magnetic field concentrations.  While the bulk radial proton ('solar wind') speed here ranges from 200-400 km/s, these measurements are akin to the 'microstream' structures measured previously in the fast solar wind \citep{1995JGR...10023389N}.  However these intervals are clearly pressure-balanced, hence spatially stable, and show enhanced alpha particle abundance, hot proton beams, energetic particles, and depressed electron temperature - all suggestive of fast wind-like sources.  The alpha particles are heated to $T_\alpha \ge $ 8$T_p$ and the temperature anisotropies of the alpha particles and beam protons suggest a common origin and/or evolution.  The ballistically-mapped longitudinal source structure and the presence of suprathermal ions suggests a source at the edge of magnetic funnels associated with interchange reconnection.  A narrow ($\sim$1$^\circ$) leading edge shows enhanced proton parallel heating, similar to measurements from Helios at the leading edge of high speed streams \citep{1982JGR....87...35M}.

We present a schematic scenario of the source suggesting that PSP is transiting through the coronal extension of magnetic funnels \citep{10.1007/bf00148082,1976RSPTA.281..339G,1986SoPh..105...35D, 2005Sci...308..519T} and possibly plumes  \citep{Wilhelm2011, Poletto2015} associated with the photospheric network magnetic field, where the Alfv\'enic switchbacks are generated.  The underlying magnetic configuration should be favorable to interchange magnetic reconnection between adjacent funnels and/or closed loop structure above the photosphere, as developed in the furnace solar wind model of \citet{1999SSRv...87...25A}.  Reconnection would also explain the energetic ions and the inherent intermittent nature of the switchbacks.  Although we highlight the impact of processes and structure in the low solar atmosphere, some properties of SBs may develop or be further amplified by {\em in situ} evolution in the solar wind, which increases $\delta B_R/B_0$ and can lead to abrupt rotations of B \citep[e.g.][]{2020ApJ...891L...2S,2021arXiv210109529S,2021arXiv210408321M}.



\section{Parker Solar Probe Measurements and  context}
We use magnetic field and electron density measurements from the FIELDS instrument \citep{2016SSRv..204...49B}, plasma ion and electron measurements from the Solar Wind 
Electrons Alphas and Protons (SWEAP) instrument suite \citep{2016SSRv..204..131K}, and energetic ion measurements from the Integrated Science Investigation of the Sun (ISOIS) suite \citep{2016SSRv..204..187M} 
on the NASA Parker Solar Probe (PSP) spacecraft \citep{2016SSRv..204....7F}.  
Measurements are made near perihelion of PSP Encounter 06 from September 27-28, 2020 when the spacecraft 
was between 20.4-22.7 solar radii ($R_S$) from the Sun center and 232$^\circ$-270$^\circ$ heliographic (HG) longitude; the spacecraft trajectory 
dips south of the ecliptic near perihelion and was below -3$^\circ$ HG latitude during this interval and stayed below the heliospheric current sheet (HCS).  Perihelion at 20.39 $R_S$ occurred at 09:16 UT on September 27, 2020.   
The spacecraft tangential (Keplerian) speed in HG coordinates ranges from 87 km/s at the start of the interval to 68 km/s at the end, passing through the maximum 89 km/s at perihelion.
The spacecraft radial speed ranges from -11 km/s at the start of the interval to 39 km/s at the end, passing through the 0 km/s at perihelion.  The measurements we analyze here are primarily in the outbound leg of the orbit.

\begin{figure*}[h]
\center{\includegraphics[width=17cm, height=11cm]{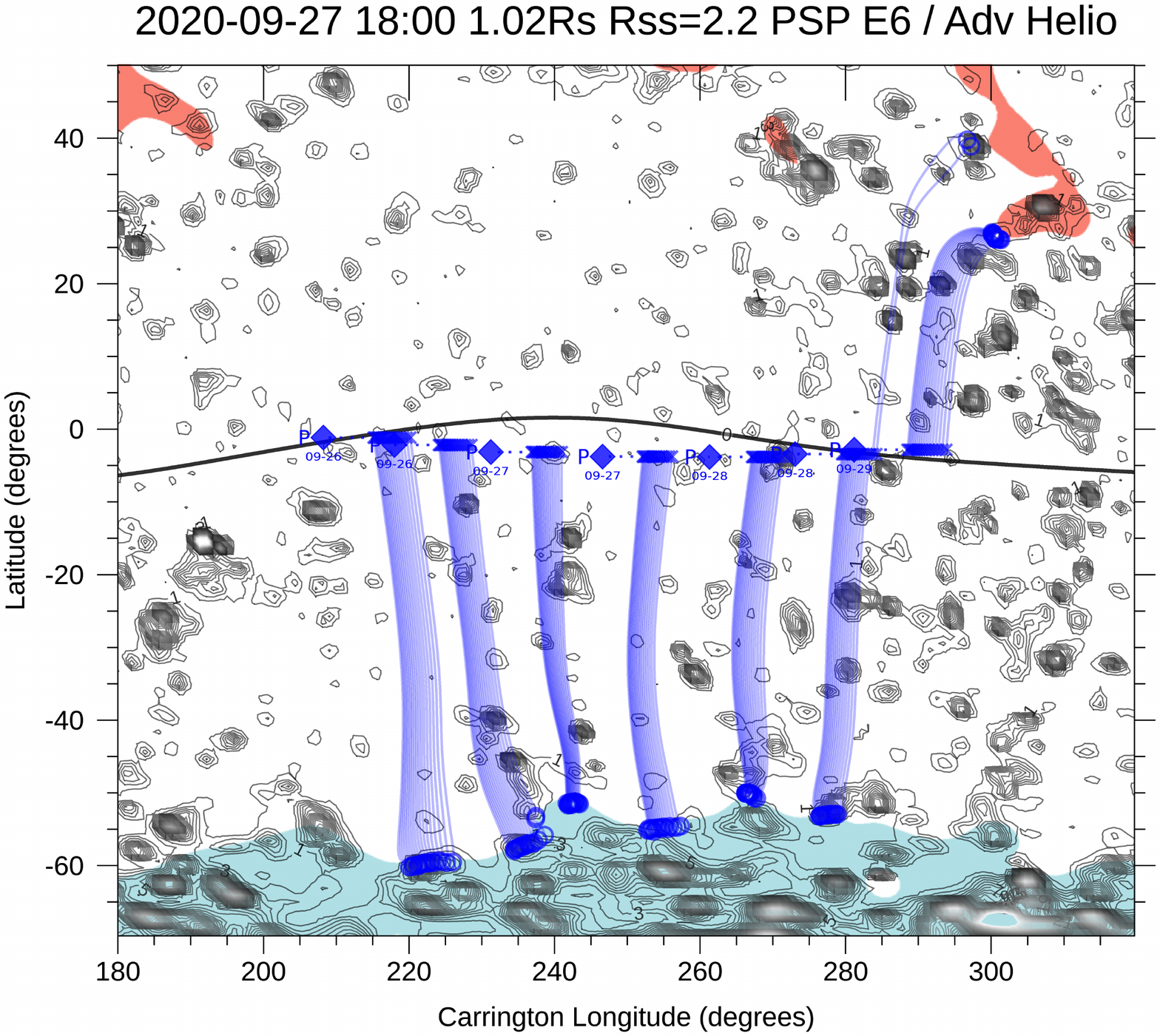}}
\caption{Magnetic field connectivity during PSP Encounter 06. The perihelion loop is seen between 210$^\circ$ and 300$^\circ$ Carrington (heliographic) longitude. A PFSS model (described in the text) maps  PSP magnetic connection (in 12 hour intervals) from the source surface at 2.2 Rs to the height 1.02 Rs above the photosphere.  A solid black line shows the model neutral line on the source surface. Black contours indicate magnetic field pressure at 1.02 Rs ($\sim$ 14 Mm) showing the spatial scales associated with network magnetic field and supergranulation cell boundaries. During the interval studied here, the source surface was connected to the southern coronal hole boundary near -60$^\circ$ Carrington latitude.  The ballistic projection of the PSP trajectory (blue diamonds) on the source surface (blue crosses) and down to the solar wind source regions (blue circles) calculated for the height R = 1.02 $R_S$ and measured {\em in situ} solar wind speed $\pm$ 80 km/s. Open magnetic field regions are shown in blue (negative) and pink (positive). The $B^2$ map is calculated for 27 Sep 2020, different PSP positions along the trajectory shown for selected days are displayed under the blue diamonds with a time cadence of 12 hours during Sep 26-29, 2020.}
\label{fig:pfss}
\end{figure*}

Figure \ref{fig:pfss} shows the magnetic connectivity of the PSP spacecraft footpoints from the source surface at $R_{SS}$ = 2.2 $R_S$ mapped down the solar surface, near the low latitude boundary of a coronal hole (CH) feature at -60$^\circ$ HG latitude.  The mapping uses a Potential Field Source Surface (PFSS) model 
\citep{1969SoPh....6..442S,1969SoPh....9..131A,1984PhDT.........5H} as implemented by \cite{2003SoPh..212..165S}.
   PFSS has proved to be remarkably robust during early PSP (solar minimum) orbits \citep{2019Natur.576..237B,2020ApJS..246...23B,2020ApJS..246...54P}.  
  As a lower boundary condition, the PFSS model incorporates magnetic field maps produced by an evolving surface-flux transport model based on magnetic fields observed by the Helioseismic and Magnetic Imager (HMI) \citep{2012SoPh..275..207S,2012SoPh..275..229S} on the Solar Dynamics Observatory (SDO). The model evolves these fields in accordance with empirical prescriptions for differential rotation, meridional flows, and convective dispersal processes.  PFSS models are parametrized by a radial distance at which all magnetic field lines become open and radial: the source surface (SS) height.  This is typically set at $R_{SS}$ = 2.5 $R_S$, but is a free parameter of the model and we choose $R_{SS}$ = 2.2 $R_S$ to be compatible with {\em in situ} magnetic field measurements by PSP and Solar Orbiter for this time interval \citep{2021ApJ...912L..21T}.

 Black contours in Figure \ref{fig:pfss} indicate magnetic field pressure at 1.02 $R_S$ ($\sim$14 Mm altitude above the photosphere) and reveal a network of stronger magnetic field concentrations, which can be inferred up to $\sim$1.04 $R_S$, or 28-30 Mm and then quickly dissipate higher in the corona due to magnetic field line expansion. This low altitude, about 10 times the height of the chromosphere, is found to be critical in filament channel formation and dynamics \citep{2009ASPC..415..196P} and is comparable to a typical supergranulation cell diameter. 
The boundary of the southern coronal hole along which PSP footpoints were moving has a sequence of these magnetic concentrations (nodes) with much weaker field between; we estimate from PFSS that PSP crossed 6-8 strong field concentrations during the Sep 26-29 interval studied here.   Note that this PSP perihelion occurred behind the limb, so that there were no current magnetogram data available for the longitudes below the spacecraft.  Hence there is no expectation of a one-to-one correspondence between individual network field concentrations seen in the PFSS, which evolve on timescales of several hours \citep{10.12942/lrsp-2010-2}, and our {\em in situ} measurements.



Proton beam and core parameters are obtained by fitting drifting bi-Maxwellians to the SWEAP/SPAN-Ion proton spectra. The proton beam is constrained to lie along the magnetic field direction relative to the core velocity.  For the alpha parameters, 4 successive spectra were first summed together to obtain better statistics. The SPAN-Ion alpha channel contains a small ($\sim 2\%$) proton contamination. This is compensated for by taking the previously fitted proton parameters and scaling the total density to represent the spurious protons in the alpha channel. This scaled down function, as well as a single bi-Maxwellian to represent the alpha particles, are then fitted to the alpha channel counts spectra and the alpha core parameters obtained. 
The exact scaling factor from proton channel to alpha channel is a free parameter in the fit, and it was confirmed that there was no energy or angle dependence to the proton contamination, so that an overall simple scaling was sufficient.  Distribution functions are fitted only when the solar wind distribution is within the SPANi field-of-view \citep[e.g.][]{2021A&A...650L...1W}.  For a more detailed discussion of SPAN-Ion fitting procedures, see \citep{2021A&A...650A..17F}.
SWEAP/SPAN-E measurements \citep{2016SSRv..204..131K,2020ApJS..246...74W} are used to determine electron core and strahl parameters by a combination of fitting and partial moment calculations, as described in \cite{2020ApJS..246...22H}.  The electron core component is fitted to a drifting bi-Maxwellian function and the halo component, where present at measurable levels, to an isotropic non-drifting Maxwellian function.  Partial moments of the strahl are computed by integrating over the residual of the measured distribution with respect to the core (and halo, if present), for the portion of the distribution in the strahl direction, within 45$^\circ$ of the magnetic field, and for velocities greater than two thermal speeds.


\begin{figure*}[h]
\includegraphics[width=18.5cm, trim=3cm 3cm 1.5cm 1cm, clip=true]{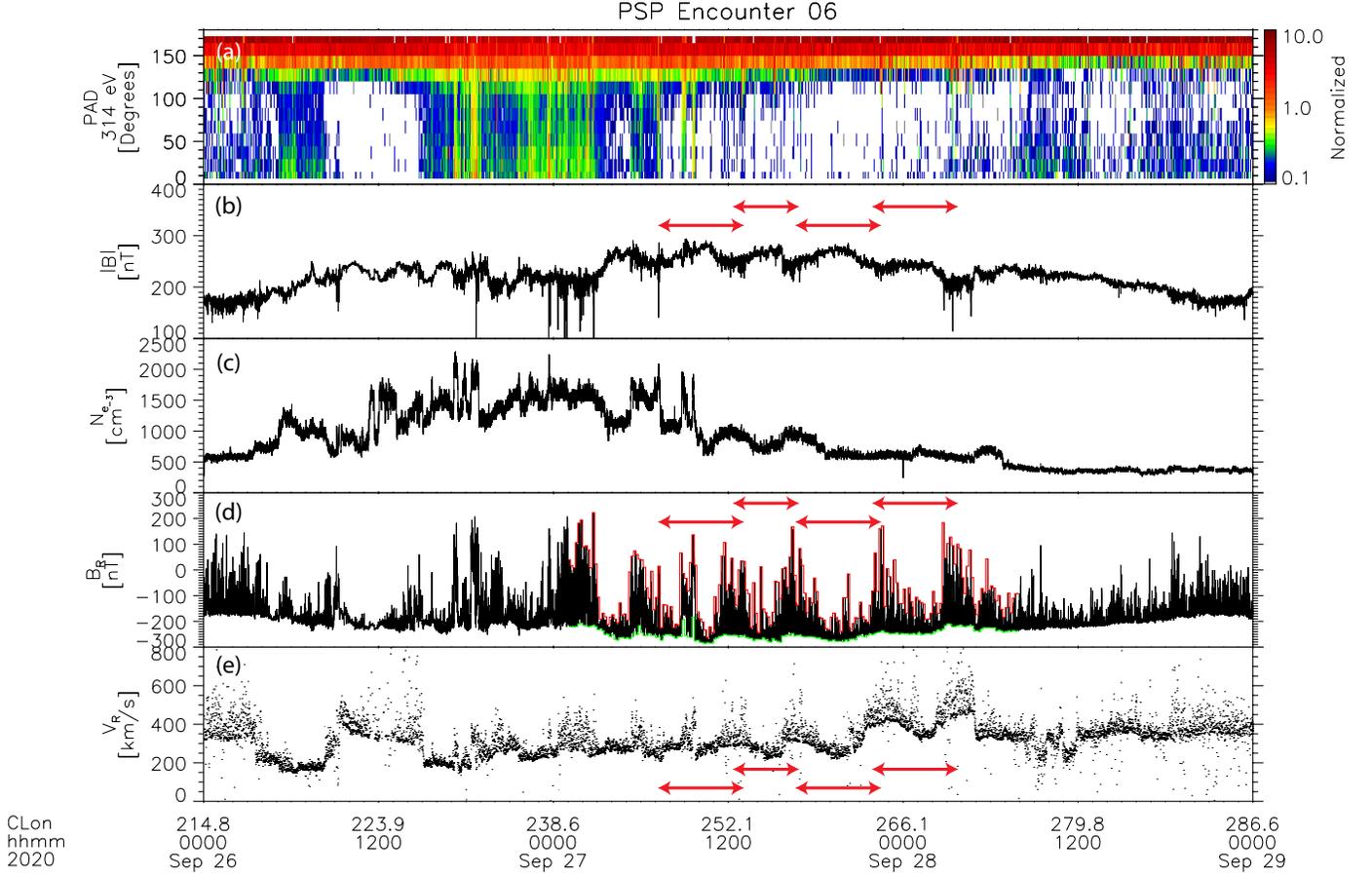}
\caption{Measurements of the suprathermal (314 eV) electron PAD (panel a), magnetic field magnitude (b), plasma density (c), radial magnetic field $B_R$ (d), and 
radial proton core velocity $V_R$ (e) during the PSP Encounter 06 perihelion.   Red horizontal arrows highlight the magnetic field compressions/depletions, velocity modulations and the accompanying modulations in the switchback occurrence and amplitude.  The red and green bars in panel (d) are the 10-minute minimum and maximum values of $B_R$.  The ballistically-mapped Carrington longitude is shown at the bottom:  these modulations occur on angular scales of a few degrees.  The outward electron strahl flux implies that the field geometry is uniform outward.}
\label{fig:overview}
\end{figure*}

Magnetic field and plasma measurements from PSP Encounter 06 are shown in Figure \ref{fig:overview}.  A normalized (to 90$^\circ$) pitch angle distribution (PAD) of 314 eV suprathermal (strahl) electrons after $\sim$ 01:00 on September 27, 2020 show primarily outward flux along $-B_R$ indicating a unipolar magnetic field geometry below and consistent with the PSP trajectory being below the HCS as implied in Figure \ref{fig:pfss}.  Periodic modulations of the magnetic field magnitude $|B|$ (panel b) and radial proton core velocity $V_R$ (panel e) are emphasized using red arrows.  Panel (d) shows the radial component of the magnetic field $B_R$ and the minimum (green) and maximum (red) value envelopes in 10 minute intervals.  The $B_R$ dynamics are dominated by magnetic field 'switchbacks' \citep{2019Natur.576..237B,2020ApJS..246...45H} that here are clearly modulated in occurrence and amplitude by the structure in $|B|$ and $V_R$.  Small plasma density enhancements are also measured for some of these intervals (panel c).  It has been reported previously that the Alfvenic magnetic field switchbacks are interspersed in regions of 'quiet radial flow' \citep{2019Natur.576..237B,2020ApJS..246...45H}; here at 20 $R_S$ they are clearly modulated periodically and in correspondence with field amplitude and proton radial flow.  Along the bottom of Figure \ref{fig:overview} the spacecraft position in Carrington (HG) longitude is shown, as mapped ballistically using the measured proton speed \citep[e.g.][]{1973SoPh...33..241N,2020ApJS..246...23B}.  The time interval of the modulated features (red arrows) corresponds to Carrington longitudes of $\sim$ 250$^\circ$ - 270$^\circ$ in Figure \ref{fig:pfss}.  The modulations of switchbacks and plasma parameters occur on angular scales of a few degrees, similar to the network magnetic field and supergranulation scales on the Sun as in Figure \ref{fig:pfss}.

\begin{figure}[h]
\centering{\includegraphics[width=16
cm, height=19cm, trim=2cm 2.5cm 2cm 2.5cm, clip=true]{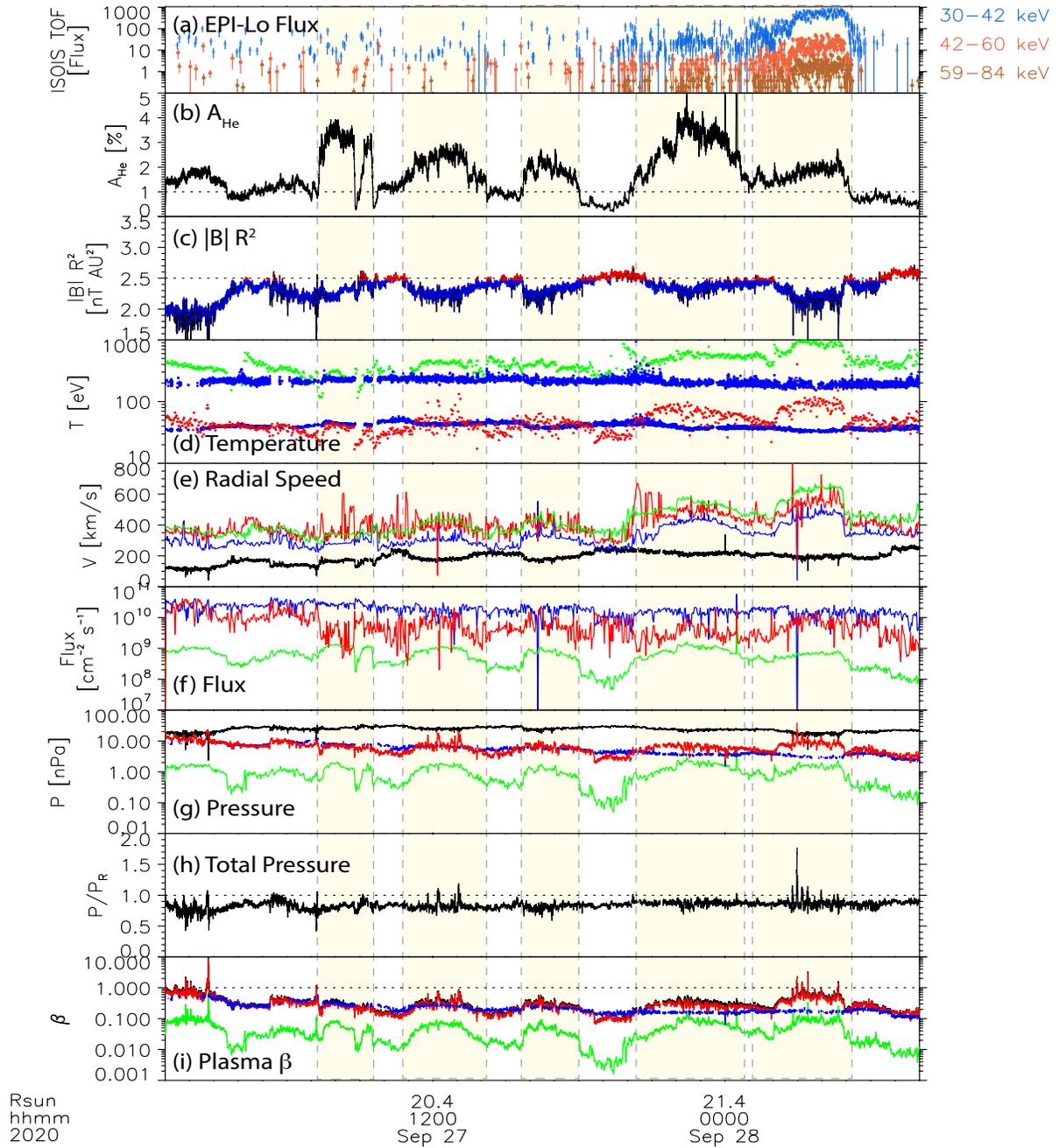}}
\caption{Measurements of the suprathermal ion number flux (panel a), alpha particle abundance $A_{He}$ (b), scaled magnetic field magitude $|B| R^2$, 
electron core and strahl temperatures (blue), core proton temperature (red), and alpha temperature (green) in panel (d),  radial 
velocities (core proton = blue, alpha = green, proton beam = red) and Alfven speed (black) in panel (e).   Number flux of core protons (blue), alphas (green) and beam protons (red) is in panel (f) and plasma pressures (core proton = red, 
alpha = green, electron = blue) and magnetic field energy density (black) in panel (g).  Panel (h) is the total pressure, plasma and magnetic, scaled 
by its fitted radial scaling $p \propto r^{-3.17}$.  The bottom panel shows plasma beta (red = core proton, green = alpha, blue = core electron).  The intervals of depressed field (panel b), which correspond to enhanced switchbacks, have enhancements in alpha abundance $A_{He}$ consistent with fast solar wind values.  Likewise, the intervals of enhanced switchbacks have higher speed radial wind speeds (panel d) with (core) Mach $\sim$ 1 near the 
edges but higher Mach flows in the center.  Ion temperatures are higher within these structures and electron temperatures are slightly lower.  Alpha abundance, speed and temperature measurements suggest that the switchback intervals are more like 'fast wind'.
Panel (g) demonstrates that the set of structures is in local pressure-balance consistent with spatially 
stable structure on the spacecraft transit timescale.}
\label{fig:pressure}
\end{figure}

Figure \ref{fig:pressure} highlights the interval of modulated switchbacks between 2020-09-27/01:00:00 and 2020-09-28/08:00.  The top panel shows ion flux measurements from ISOIS in 3 energy channels from the EPI-Lo time-of-flight (TOF) system between 30-84 keV.
An examination of the triple coincidence data shows that these TOF-only measurements are likely to be dominated by a mix of He and/or O ions and are field-aligned (anti-sunward).  Note that these suprathermal ions appear most clearly within the last two modulations (in yellow bars), which also contain higher speed flow (panel (e)), suggesting that these ions may be the suprathermal tail of the modulated solar wind distribution.
Panel (b) is the measured alpha particle abundance $A_{He} = n_\alpha/n_{total}$ showing strong modulations to relatively large values ($\sim 4\%$) associated with the switchback patches and magnetic field amplitude modulations in panel (c).   The dashed line in Panel (c) indicates the nominal heliospheric field of 2.5 nT AU$^2$ \citep{2021A&A...650A..18B} and the depressed intervals are colored blue. Light yellow vertical panels indicate intervals with enhanced $A_{He}$ and depressed $|B|$.  Alpha particle temperature $T_\alpha$ (green) and proton core temperature $T_p$ (red) are enhanced within these intervals and electron core (lower blue) and strahl (upper blue) temperatures are depressed; note that $T_\alpha/T_p \approx 4-15$ as will be discussed below.  Plasma radial velocities (panel e) are enhanced within the structures with the core proton speed (blue) approaching the Alfv\'en speed (black) on the edges and Alfv\'en Mach $M_A \approx$ 2 within.  The alpha (green) and proton beam (red) speeds are a large fraction of an Alfv\'en speed.  The inverse correlation of electron temperature to wind speed $T_e \propto 1/v_{sw}$ is well known and observed quite universally during PSP solar encounters \citep[e.g.][]{2020ApJS..246...62M}.

Panel (f) in Figure \ref{fig:pressure} show the number flux of core protons (blue), beam protons (red) and alpha particles (green).  The core proton flux is relatively steady, while the proton beam and alpha particle flux are more modulated by the plasma structure; the proton beam abundance (not shown) is not so clearly modulated, rather the proton beam speed is enhanced within the high alpha abundance structures.  Magnetic field pressure (black) and plasma pressure are shown on Panel (g), with proton (red), core electron (blue) and alpha (green) all showing anti-correlation with the magnetic field pressure.  This is best seen in Panel (h) where the total pressure (magnetic plus plasma) is shown, normalized to a slowly varying factor to remove the large-scale radial trend - these structures are in local pressure balance.  Pressure-balanced structure (PBS) is seen throughout the solar wind in the inner heliosphere \citep{1990AnGeo...8..713T}, at 1 AU \citep{2016JGRA..121.5055B}, and on Ulysses \citep{1995JGR...10019893M,1996A&A...316..368M} and \cite{1999GeoRL..26.1805R} reported an association between PBS and enhanced alpha particle abundance.  Here we take this as evidence that these structures are spatial and stable over the spacecraft transit time, at least.  The lower panel shows plasma $\beta$ for proton (red), alphas (green), and electrons (blue) demonstrating the ion $\beta$ is enhanced inside the structures.  In addition to the measurements shown in Figure \ref{fig:pressure}, these intervals show modulations of the proton core temperature anisotropy $T_{\perp}/T_{\parallel}$ that often exceed the anisotropy-driven ion cyclotron instability threshold \citep[e.g.][]{2006GeoRL..33.9101H}, relatively large proton beam speeds and alpha-proton drifts (comparable to $v_A$) and depressed $T_e/T_p$ and the interval is richly populated with ion cyclotron-frequency waves \citep{2020ApJS..246...66B,2020ApJS..248....5V}.  Enhanced proton-alpha drifts within PBSs have been reported previously by \cite{2004JGRA..109.3104Y}.  The kinetic features measured here are also associated with enhanced electrostatic plasma waves and will be investigated more fully in a future study.

In summary, Figure \ref{fig:pressure} demonstrates that PSP is transiting over spatial plasma structures that are a few degrees of HG longitude across mapped to the Sun.  The structures contain hotter and faster ions (protons and alpha particles), a markedly enhanced alpha particle abundance $A_{He}$, suprathermal ions, depressed magnetic field $|B|$ and electron temperature $T_e$ and a clear increase in the amplitude and occurrence of magnetic field switchbacks.  Since the alpha abundance, and arguably the electron temperature, are frozen-in from the solar wind source in the transition region, we argue that the suprathermal ions have an origin at these altitudes and that the physics of these solar wind sources organizes the spatial and temporal distribution of the switchbacks.

\begin{figure}[h]
\centering{\includegraphics[width=18cm, height=8.5cm, trim=0cm 0cm 0cm 0cm, clip=true]{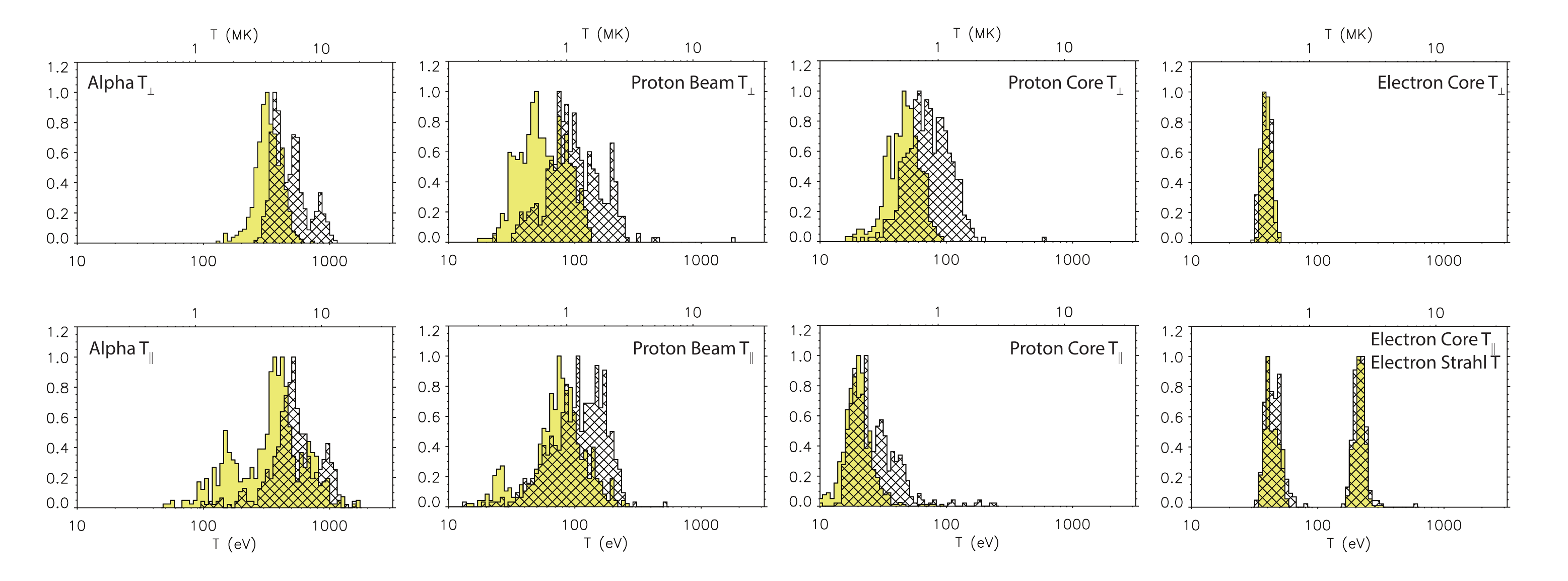}}
\caption{Distribution of temperatures (normalized with the peak to one) for alpha particles, beam protons, core protons, and core and strahl electrons, both perpendicular to the magnetic field (upper row) and parallel (lower row).  The hatched histograms are within the funnel structures and the beige histograms are outside of the funnels.  Temperatures are given in electron volts (lower axis) and MKelvin (upper axis).  Protons and alpha particles are all hotter within the funnel structures, as seen in Figure \ref{fig:pressure} as well.  Electron $T_\perp$ is similar inside and out, with electron $T_\parallel$ slightly enhanced within the funnels.  Characteristic values are given in Table \ref{table:temperatures}}
\label{fig:temperatures_all}
\end{figure}

\begin{figure}[h]
\centering{\includegraphics[width=7cm, height=11cm, trim=0cm 0cm 0cm 0cm, clip=true]{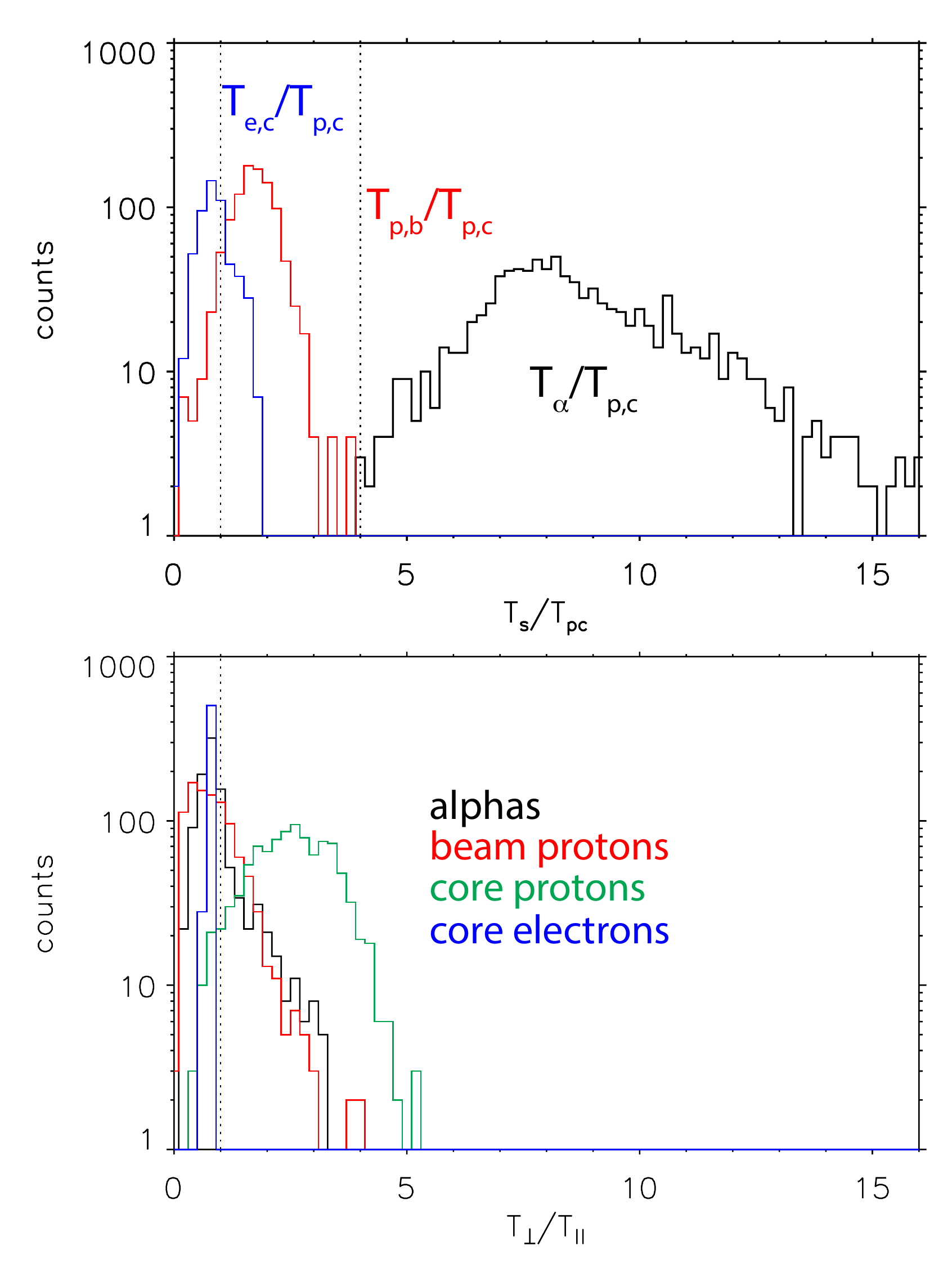}}
\caption{Distribution of temperatures (upper panel) and temperature anisotropy (low panel) for alpha particles (black), beam protons (red), core protons (green) and core electrons (blue).  The alpha particle total temperature shows clear evidence of heating above mass ratio $T_{\alpha}/T_{p,c} > m_{\alpha}/m_{p} = 4$ and peaks near $T_{\alpha}/T_{p,c} \approx$ 8.  The distribution of proton beam temperatures peaks at near $T_{p,b}/T_{p,c} \approx$ 2 and the core electron population is slightly cooler than the protons $T_{e,c}/T_{p,c} <$ 1.  While the core protons are strongly anisotropic $T_{\perp}/T_{\parallel} \ge$ 2, the alpha particles, beam protons, and electrons all show slightly enhanced parallel temperatures $T_{\perp}/T_{\parallel} \lesssim$ 1.  Note that the distributions of alpha and beam proton temperature anisotropies are very similar suggesting a common heating and/or evolution mechanism.}
\label{fig:temperatures}
\end{figure}

Histograms of ion and electron temperatures are shown in Figure \ref{fig:temperatures_all}, normalized to unity maximum and characteristic values are collected in Table \ref{table:temperatures}.  The cross-hatched histograms are accumulated within the high-alpha structures and the beige histograms are from outside the structures.  The proton and alpha populations are generally hotter within the magnetic structures while both core and strahl electrons are slightly cooler, reminiscent of fast solar wind.  Variations from structure-to-structure mask some of the trends in the unnormalized temperature that become more apparent in the dimensionless ratios in Figure \ref{fig:temperatures}. 

\begin{table}[htp]
\begin{center}
\begin{tabular}{l|ccc|ccc}
  &  &$\tt{inside/outside}$&  & & $\tt{inside/outside}$ &   \\
Species & $\langle$ T$_\perp \rangle$ &$Mo($T$_\perp)$& $ \sigma($T$_\perp$) & $\langle$ T$_\parallel \rangle$ & $Mo($T$_\parallel)$ & $ \sigma($T$_\parallel$)    \\
\hline
He$^{++}$  & 501/335 & 360/305 & 171/79 & 570/400 & 464/340 & 247/241  \\
Beam H$^+$   & 114/60 & 75/47 & 96/25 & 117/81 & 105/74 & 54/40  \\
Core H$^+$  & 78/47 & 58/40 & 37/13 & 30/21 & 17/21 & 23/8  \\
Core e$^-$  & 39/40 & 36/39 & 6.2/6.3 & 45/43 & 39/41 & 6.7/6.6  \\
Strahl e$^-$  & n/a & n/a & n/a & 217/215 & 216/220 & 15/15  \\
\hline
\end{tabular} 
\caption{Average, mode, and standard deviation of the distribution of T$_\perp$ and T$_\parallel$ in units of eV, inside and outside of the funnel structures (inside/outside).}
\end{center}
\label{table:temperatures}
\end{table}

\cite{stansby2019} compared Helios measurements of alpha and proton temperatures down to 0.3 AU to expectations from double-adiabatic theory \citep{Chew.PRSA.1956} (or CGL).  They found that $T_{\alpha, \parallel}$ decreases faster with radial distance than predicted by CGL, while $T_{\alpha, \perp}$ decreases more slowly, consistent with isotropization associated with temperature anisotropy instabilities \citep{Hellinger.GeophysicalResearchLetters.2006,10.1103/physrevlett.103.211101,Maruca.TheAstrophysicalJournal.2012}.  Our average values within the funnel intervals at $\sim$ 20$R_S$ ($\sim$ 0.1 AU) of $\langle T_{\alpha, \parallel}\rangle \approx$ 570 eV (6.6 MK), $\langle T_{\alpha, \perp} \rangle \approx$ 501 eV (5.8 MK), $\langle T_{p, \parallel}\rangle \approx$ 30 eV (0.35 MK), and $\langle T_{p, \perp} \rangle \approx$ 78 eV (0.9 MK) are consistent with the trends in Figure 1 of \cite{stansby2019} and $\langle T_{\alpha, \perp} \rangle$ seems to trend with their CGL curve.
In Figure \ref{fig:temperatures} we show the distribution of relative (to proton core) temperatures (top panel) and the temperature anisotropy $T_\perp/T_\parallel$ for each population, over the interval.  Notably, the alpha particles are heated substantially relative to the core protons with $T_\alpha$ peaking near $T_\alpha \sim$ 8 $T_{p,c}$, well above, even twice, the mass-proportional heating $T_\alpha \sim$ 4 $T_{p,c}$ rate.  More-than-mass proportional heating has been measured in the inner heliosphere previously \citep{1982JGR....87...35M, 2012JGRA..117.0M02G}, and at 1 au \citep{Kasper.ApJ.2017}
 and here we identify it with discrete solar wind sources and report, we believe, one of the largest ever heating fractions $T_\alpha \gtrsim$ 10 $T_{p,c}$ for thermal alpha particles in the solar wind.  Notably, there is no thermalized population $T_\alpha \sim T_{p}$.   Indeed, \cite{2013PhRvL.111x1101M} used measurements of $T_\alpha/T_{p}$ at 1 AU and a model of collisional evolution to infer large temperature ratios in the inner heliosphere; their model removed the 1 AU isotropic population $T_\alpha \sim T_{p}$ and the distribution was predicted to peak at $T_\alpha \sim$ 5.4 $T_{p}$ at 0.1 AU ($\sim$ 21 $R_S$).  If robust and general, our results imply source of isotropization in addition to collisions, presumably wave-particle effects \citep[e.g.][]{2018PhRvL.120t5102K}.  Note that alpha-to-proton temperature ratios of $\sim$10 were predicted at 15 $R_S$ by \cite{Chandran.ApJ.2010} based on stochastic ion heating by low frequency MHD/KAW turbulence.

The proton beam population is heated to $T_{p,b} \sim 2 ~T_{p,c}$, while the core electrons are cooler than the core protons $T_{e,c} \lesssim T_{p,c}$ as observed for faster solar wind sources \citep{1989JGR....94.6893M, 1998A&A...336L..90D,2020ApJS..246...62M}.  The lower panel of Figure \ref{fig:temperatures} shows the temperature anisotropy $T_{\perp}/T_{\parallel}$ of the core electron (blue), core proton (green), beam proton (red) and alpha particle (black) populations.  The core proton population is notably more anistropic; the proton beam and alpha particle populations show similar distributions of temperature anisotropy.  This may be a result of a common heating or energization mechanism, or rather may indicate similar physics during radial evolution.  Since the beam protons and alpha particle stream ahead of the core protons, these populations will interact differently with Alfv\'enic fluctuations propagating in the wave frame with respect to the bulk (core proton) solar wind.

\begin{figure}[h]
\centering{\includegraphics[width=15cm, height=6cm, trim=0cm 0cm 0cm 0cm, clip=true]{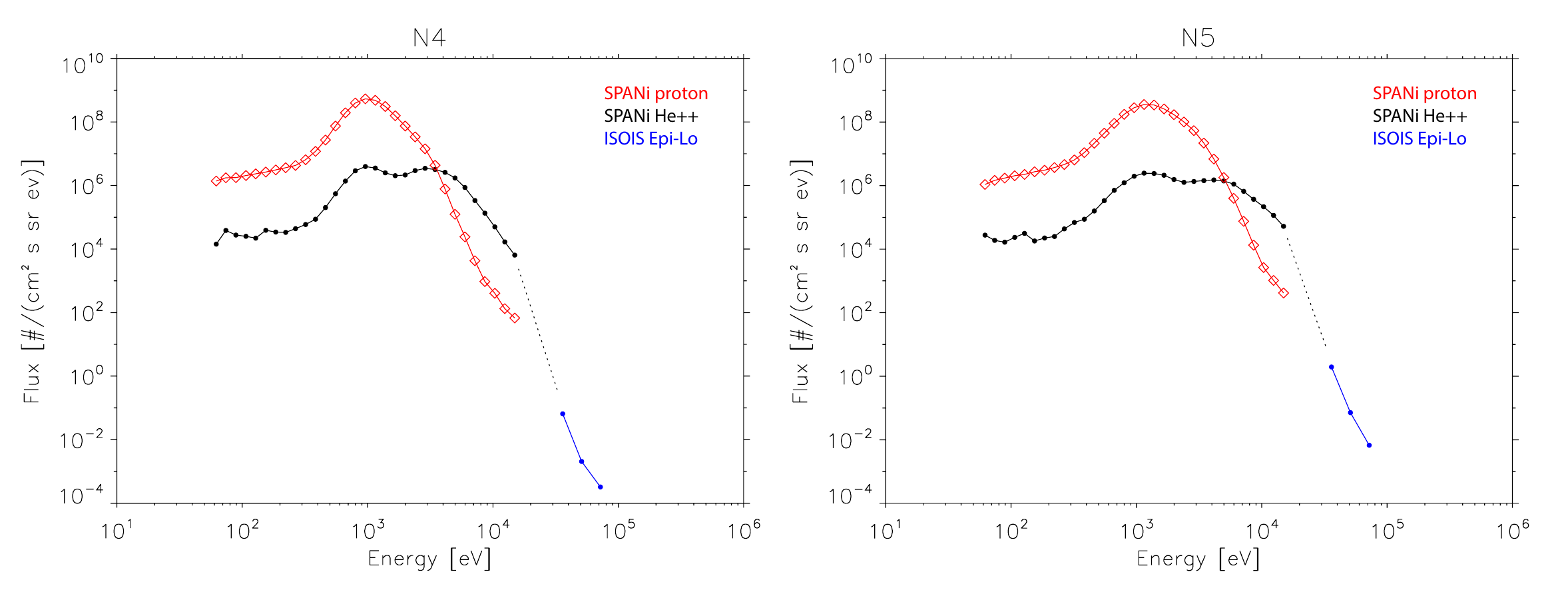}}
\caption{Energy spectra in units of differential number flux during the interval within the last two yellow-highlighted funnel structures of Figure \ref{fig:HGcoords} (labeled N4 and N5 there).  A dotted line with spectral 
index $\sim -11$ extends between the SPANi alpha and ISOIS measurements; we suggest that the ISOIS ions are the suprathermal tail of the solar wind alpha spectrum. }
\label{fig:spectra}
\end{figure}

\begin{figure}[h]
\centering
   \includegraphics[width=14cm, height=19cm, trim=1cm 1.5cm 2cm 0.5cm, clip=true]{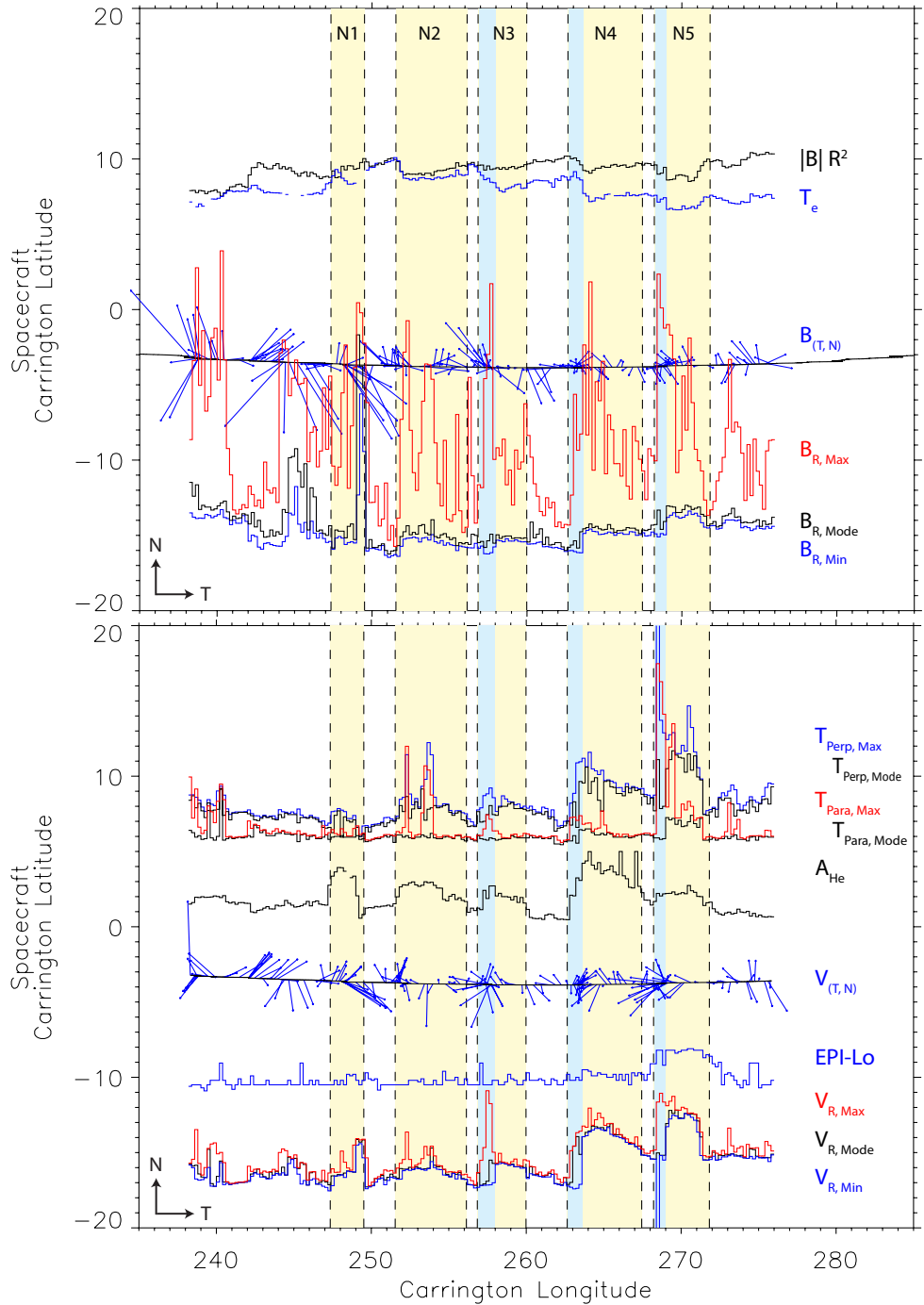}
\caption{Magnetic field (upper panel) and proton velocity (lower panel) vectors in the T-N plane and centered on the 
spacecraft position in latitude and ballistically-mapped Carrington longitude.  Field magnitude $|B|~R^2$, electron core temperature $T_e$ and radial field $B_R$ (upper panel) and proton $T_\perp$ and $T_\parallel$, $A_{He}$, and radial core proton speed $V_R$ (lower panel) are overplotted in arbitrary units.  Maximum, minimum, and mode values are shown for $B_R$ and $V_R$ and maximum and mode for the proton temperatures.  Data are accumulated in bins of 0.2$^\circ$ longitude.
Yellow vertical bars show intervals (labeled N1-N5) of enhanced $A_{He}$ within the PBS and blue vertical bars indicate subintervals with enhanced temperature and radial flow.  The T-N plane field and velocity vectors show non-radial structure 
near the boundaries of the $A_{He}$ enhancements.  $B_{R, Max}$ (red bars, upper panel) indicates the distribution of switchbacks; occurrence and amplitude is organized by the PBS and peaks near, but somewhat within, the leading edge at the source on the Sun.  Intervals of enhanced proton $T_\parallel$, radial speed, and electron temperature $T_e$ occur within $\sim1^\circ$ of the (spacecraft-frame) leading edge of structures N3-N5.}
   \label{fig:HGcoords}
\end{figure}

 Figure \ref{fig:spectra} shows energy spectra of protons and alpha particles measured by SPANi compared with the ISOIS EPI-Lo spectrum; these spectra are in units of differential number flux $j$ and averaged over the intervals 2020-09-27/21:28:00 - 2020-09-28/00:10:00 (left) and 2020-09-28/02:44:30 - 2020-09-28/04:55:30 (right), corresponding to the final two yellow boxes in Figure \ref{fig:pressure} and labeled N4 and N5 in Figure \ref{fig:HGcoords}.  The ISOIS suprathermal spectrum is shown assuming a spectrum of He$^{++}$ (as described above); if the ISOIS spectrum is oxygen the fluxes will be a factor of 4 higher.  While suprathermal He$^{++}$ is observed routinely in the ecliptic, slow solar wind \citep[e.g.][]{Collier.GeophysicalResearchLetters.1996} most usually associated with shocks or CIRS (or pickup ions), Ulysses measurements show a negligible ambient suprathermal He$^{++}$ population in the fast solar wind \citep{Gloeckler.SpaceScienceReviews.1998}.  Our measurements here, with the striking correlation in Figure \ref{fig:pressure} and the plausible continuity of the spectrum in Figure \ref{fig:spectra}, suggest that the EPI-Lo ion measurements are the tail of a very steep alpha particle spectrum out to 85 keV.  A dotted line in Figure \ref{fig:spectra} connects the SPANi and EPI-Lo spectra and has a spectral slope $j \sim E^{-11}$.  There is no notable radio emission nor flaring activity at this time.

To infer the spatial structure at the origin, we project the PSP spacecraft location in HG longitude down to the solar surface using a ballistic mapping \citep{1973SoPh...33..241N,2020ApJS..246...23B} in Figure \ref{fig:HGcoords}.   The $T,N$-plane magnetic field (upper panel) and proton velocity (lower panel) vectors are computed as 10-minute statistical mode values, in an attempt to remove the dominant and rapidly varying switchback fields.  A time series plot (not shown) suggests that this is largely successful and the vectors shown here represent larger scale deviations of the underlying field and flow.  Where no vector is apparent, the flow/field is primarily radial.  The vectors are centered on the spacecraft coordinates - note that the spacecraft latitude is not corrected for ballistics or footpoint location and as shown in Figure \ref{fig:pfss} the local magnetic field footpoint ultimately maps to southern HG latitudes of around -60$^\circ$.  The time series measurements of the plasma and magnetic field are overplotted in their ballistically-propagated HG longitude coordinates (and in arbitrary units here).  The result of the ballistic backprojection shows that the sources are highly-structured and steepened at the leading edge; the smooth profile at PSP altitudes is a result of time-of-flight.   The intervals with enhanced $A_{He}$ are colored yellow and labels N1-N5.  The switchback intervals (red bars $B_{R, Max}$ and $V_{R, Max}$) cluster near the leading edge or just within.  Magnetic field intensity $|B|$ and electron temperature $T_e$ are just inside this boundary; we will suggest below that this minimum in magnetic field and electron temperature correspond to the central region of an asymmetric magnetic structure.

It is interesting to note that for the last 3 events here (N3-N5), there appears to be a small ($\sim$ 1$^\circ$) leading edge (in the spacecraft frame) interval for which the maximum radial field and speed ($B_{R, Max}$ and $V_{R, Max}$) leads the step in the minimum values $B_{R, Min}$ and $V_{R, Min}$.  These intervals (colored light blue in Figure \ref{fig:HGcoords}) also correspond to decreasing trends in $|B|$ and $T_e$ and enhanced proton core temperatures $T_{\parallel}$ and $T_{\perp}$.  Similar structure was measured at stream interface boundaries with instruments on the Helios spacecraft in the inner heliosphere \citep{1982JGR....87...35M}. 

\begin{figure}[h]
\centering
   \includegraphics[width=19cm, height=10cm, trim=2.5cm 2cm 1cm 0.5cm, clip=true]{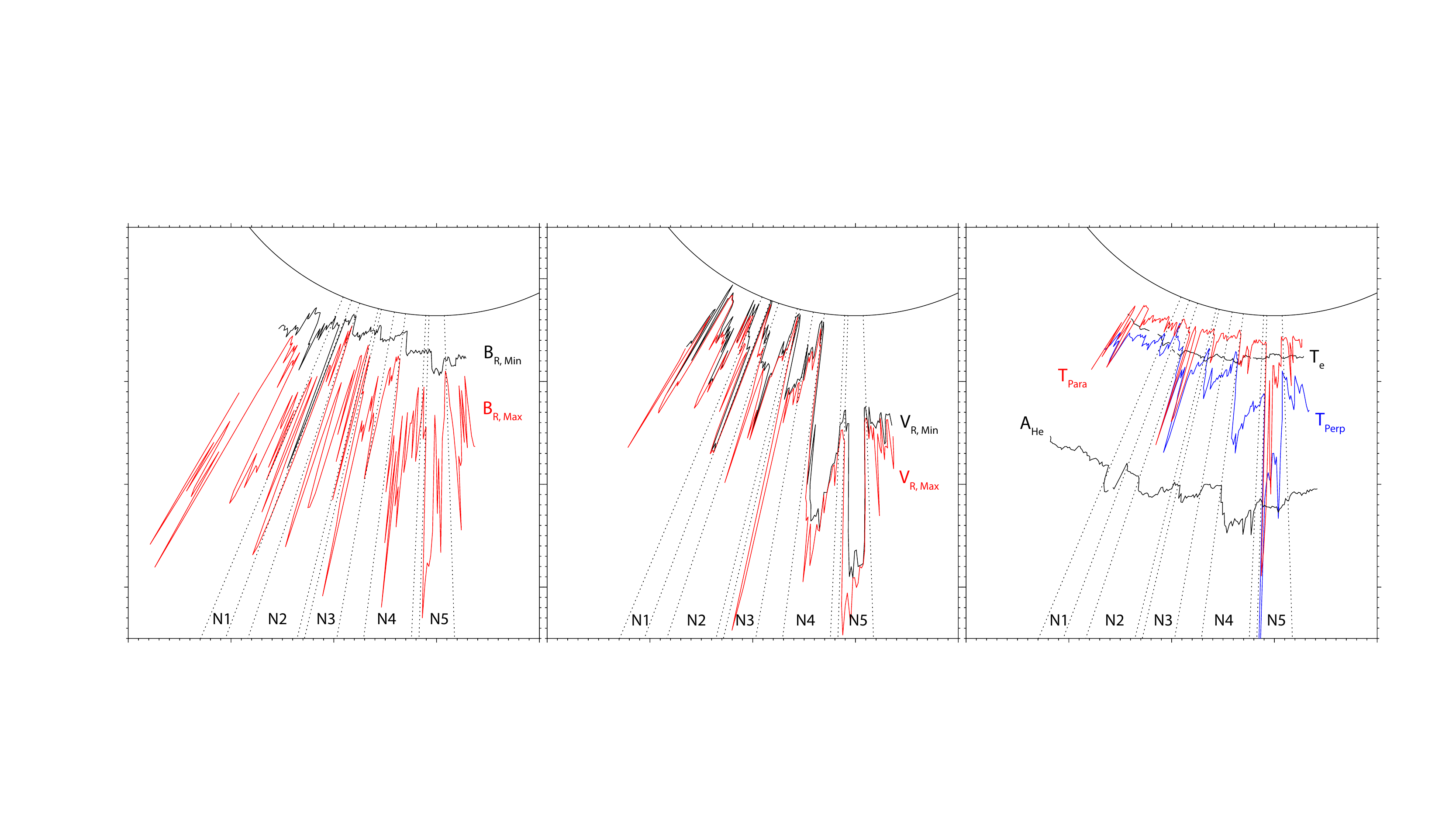}
\caption{Polar representation of the data from Figure \ref{fig:HGcoords} with the noted parameters in arbitrary units.  This shows clearly the longitudinal structure and angular scale
 of the sources.  The solar rotation is counter-clockwise. The structures show steep leading edge features in magnetic field intensity, radial proton speed, and proton $T_{\perp}$; proton $T_{\parallel}$ is peaked sharply near the leading edge of events N2-N5.}
   \label{fig:HGpolar}
\end{figure}

Figure \ref{fig:HGpolar} shows some data from Figure \ref{fig:HGcoords} represented in polar coordinates (i.e. in the plane of the heliographic ecliptic).  While the information here is equivalent to Figure \ref{fig:HGcoords}, the polar representation emphasizes the angular extent of these structures on the Sun and the potential relationship to the underlying photospheric structure.

\section{Interpretation and Discussion}

To summarize the observations in the previous section, our measurements show the following:

\begin{enumerate}[label=(\roman*), leftmargin=1.5cm,rightmargin=1.5cm,itemsep=0pt]
\item PSP passed through a $\sim$34 hour interval of modulated magnetic field switchbacks and ion and electron flux.  This interval corresponds to $\sim$25$^\circ$ of heliographic longitude at the Sun during which PSP was magnetically connected to a coronal hole boundary near -60$^\circ$ heliographic latitude.  The spacecraft was below 25 $R_S$.
\item The modulations correspond to angular scales of 3$^\circ$-5$^\circ$ in heliographic longitude at the Sun - similar to the supergranulation and network magnetic field structure.  PFSS-mapped footpoints show magnetic field concentrations on similar angular scales.
\item The modulated intervals are in total pressure-balance implying that the structures are spatial on the spacecraft transit timescale ($\sim$6 hours).
\item The thermal alpha particle abundance $A_{He}$ is enhanced to typical fast-wind values and the core and strahl electron temperatures are depressed within the modulations, implying solar wind sources on open field lines at the base of the corona similar to fast coronal hole wind.
\item  Proton and alpha particle radial speeds and temperatures and ion plasma $\beta$ are enhanced within the structures.  Alpha particles and beam protons are streaming ahead of the core protons and their number flux is strongly modulated by the structures.  The alpha particles are $\sim$ 5-15 $\times$ hotter than the core protons.  The alpha particles are fairly isotropic $T_\perp/T_\parallel \lesssim 1$, while the core protons have $T_\perp/T_\parallel > 1$ and are often unstable to an anisotropy-driven ion cyclotron instability.
\item  Suprathermal ions with energies up to $\sim$ 85 keV are measured within a subset of the structures with higher bulk velocities, suggesting a suprathermal tail.  There is no notable radio emission nor time-of-flight dispersion that suggests a classical flare origin of the suprathermal ions.  We suggest that PSP is transiting field lines populated by these ions.
\item  When mapped ballistically to the solar surface, the structures have a steep leading edge at lower Carrington/HG longitude.  This may be a signature of a some small differential rotation between the photosphere and the corona (photosphere moving faster) or maybe be some inherent asymmetry of the source fields at the coronal hole boundary.  The switchbacks cluster near the steep edge of the structures.  
\item  A narrow $\sim$1$^\circ$ region is found at the (spacecraft-frame) leading edge of the steepest structures that has enhanced proton and alpha temperatures $T_p$, $T_\alpha$, large differences between maximum and minimum radial field and flow, and a transition of $|B|$ and $T_e$.
\end{enumerate}

An enhanced relative abundance of alpha particles $A_{He} = n_\alpha/n_{tot}$ is known to be associated with fast solar wind, especially during solar minimum conditions \citep[e.g.][]{2012ApJ...745..162K}.  Since the alpha abundance $A_{He}$ is determined in the chromosphere or transition region our measurements suggest a direct mapping to discrete solar wind sources.  The modulation of wind speed and alpha abundance $A_{He}$ on the angular scales of solar supergranulation strongly implies that these 'microstreams' originate within the network magnetic field that is known to cluster at the boundaries of the supergranules \citep[viz.][]{10.12942/lrsp-2010-2,10.1007/s00159-014-0078-7}.  Theoretically this was suggested as the source of the fast solar wind in the ``furnace" model of  \citet{1999SSRv...87...25A}.  Open field lines containing hot plasma pass through the solar transition region and overexpand rapidly, as the high-order magnetic field falls off radially much more rapidly that the plasma pressure.  
The resulting magnetic 'funnel' structures \citep{1976RSPTA.281..339G,1977ebhs.coll..375G, 1986SoPh..105...35D, 1999SSRv...87...25A, 10.1126/science.283.5403.810,2005Sci...308..519T} then expand into pressure-balance to generate solar wind.  Our measurement of magnetic field intensity depressions within the structures is consistent with the overexpansion of magnetic funnels, as the axial field should decay as $B_R \sim 1/A$ while $B_{T,N} \sim 1/\sqrt{A}$  with $A$ the flux tube area, so that magnetic field at the center of the funnel (the axial,  more radial component) will be weaker after expansion.  The magnetic funnels expand rapidly at low altitudes and reconnection
of impinging small dipoles and emerging flux are the source of energy generating the solar wind within the larger coronal hole structure, whose speed is determined by the overall expansion at higher altitude, found to be  anti-correlated with the asymptotic solar wind speed \cite{Wang1990}; a two-step expansion process has been explored by \cite{1998SoPh..180..231S} and others.

\begin{figure}[h]
\center{\includegraphics[width=14cm]{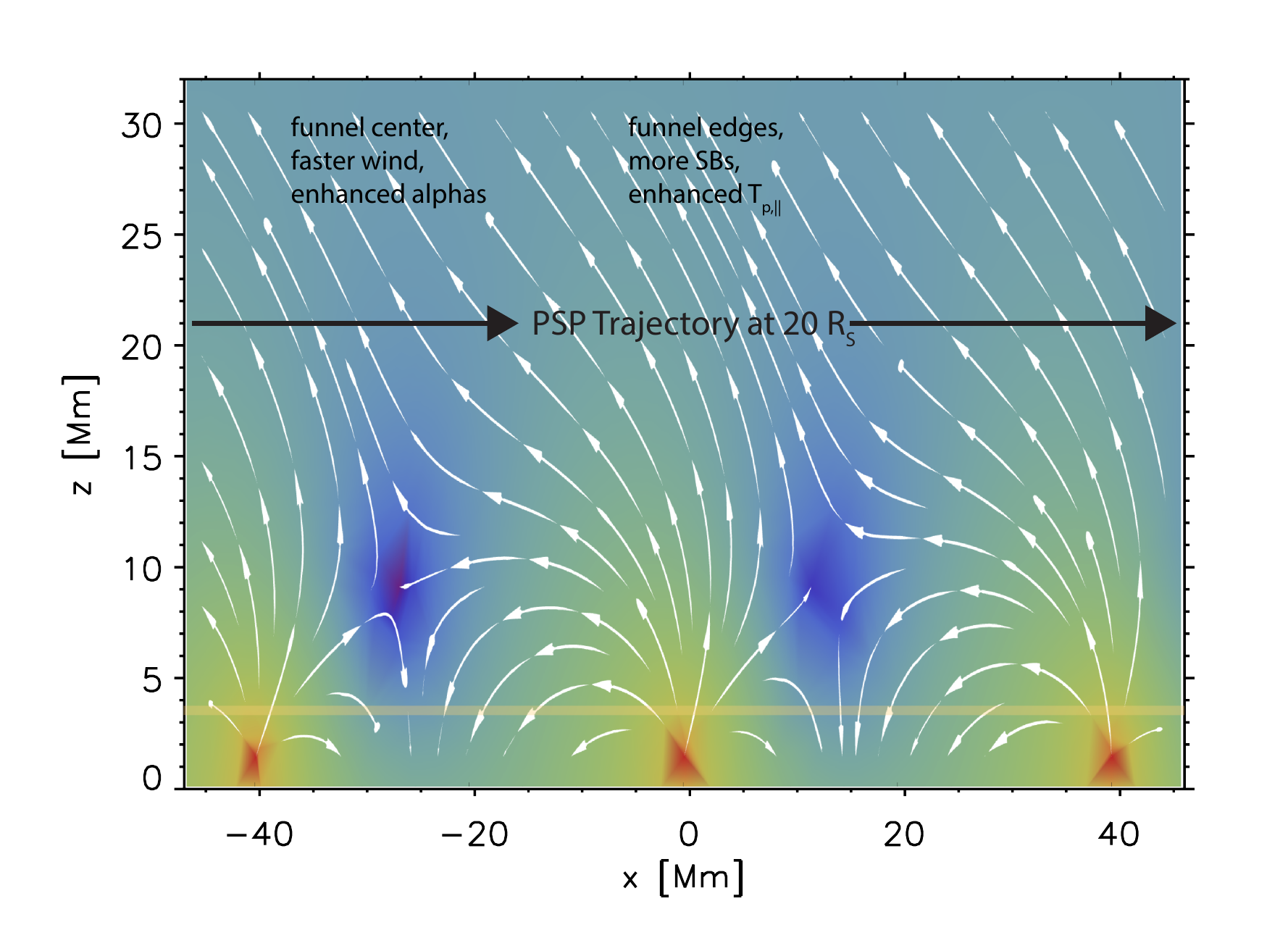}}
\caption{A schematic to illustrate the proposed configuration of magnetic field lines (white vectors) within overexpanded funnel structure and a small shear between the photosphere and corona.  The color intensity indicates field magnitude.  The bases of 
the funnels are separated by a few degrees of heliographic longitude and are generating higher $\beta$ wind with enhanced alpha particle abundance.  Patches of magnetic field switchbacks 
are localized within the funnels.  Since $B_R \sim 1/r^2$, but $B_{(T, N)} \sim 1/r$, super-radial expansion results in a depressed $|B|$ in the center of the funnel, which has spent 
more time expanding radially.  In this 2D geometry, the 'cusp' regions (darker blue) where neighboring funnels interact moves to lower altitudes with increasing shear; note that the 'cusp' intervals are $\sim$10 Mm-scale, which corresponds to $\sim$1$^\circ$.  A horizontal band indicates the transition region.
Note that the local funnel 
geometry can be expanded away by 20 $R_S$.}
\label{fig:cartoon}
\end{figure}

The association of large amplitude, nonlinear Alfvenic fluctuations (i.e. the switchbacks), enhanced bulk flow, and more than mass-proportional ion heating with the edge of the funnel structure suggests an asymmetry in the source region that is associated with the heating.  Figure \ref{fig:cartoon} is a schematic of the magnetic field structure at the source.  The field vectors here are a 2D potential model of a magnetic funnel \citep{1999SSRv...87..207H} as a set of unipolar flux concentrations and a dipole field with a shear in the -$\hat{x}$ direction; color intensity in Figure \ref{fig:cartoon} is magnetic field magnitude.

The enhancement of SBs in wind emanating from funnel boundaries clearly demonstrates the impact of structure and processes near the Sun on the spatial distribution of SBs in the solar wind. In particular, the enhanced values of $\delta B_R$ in these regions is strong evidence of preferential Poynting-flux injection near funnel boundaries in the low solar atmosphere, which could be the result of magnetic reconnection.  It is possible that the Poynting flux injection is not only stronger but also more intermittent near funnel boundaries, since reconnection is often impulsive. However, as noted previously, large-amplitude magnetic fluctuations in the solar wind naturally develop discontinuities as a consequence of spherical polarization \citep[e.g.][]{Vasquez.JGR.1996,2020ApJ...891L...2S,2021arXiv210109529S}, and thus the abrupt magnetic-field rotations observed by PSP may also originate through {\em in situ} dynamics.


The cores of the funnels are separated by 40 Mm ($\sim$3.3$^\circ$) here and already at 30 Mm altitude the field direction becomes relatively uniform, but is modulated in intensity as described above due to the superradial expansion below. The low altitude 'cusp' regions (shaded dark blue in Figure \ref{fig:cartoon}) where adjacent funnels interactions result in an x point (in 2D) with a spine-fan intersection and a dome of confined flux, where magnetic reconnection could leak heated and confined plasma outward. This asymmetry is seen in the hotter, faster leading edge of the measurements in Figure \ref{fig:HGcoords}.
  Indeed, simulations of reconnection between funnels and emerging flux show this asymmetry \citep{10.1088/0004-637x/751/2/152,10.1093/pasj/65.3.62} and the resulting outflows depend on the altitude of the reconnection region.  Here the  cusp regions move to lower altitude with increasing (applied) magnetic shear ($\hat{x}$ component); magnetic shear could be due to differential rotation between the photosphere and corona or a property of the structure of the coronal hole boundary.  
 
Magnetic reconnection may also naturally explain the energization of alpha particles (or oxygen) to 85 keV and the altitude of a reconnection site may be related to alpha particle abundance, if gravitational settling plays a role.  If the full magnetic energy of a reconnection outflow is available to energize alpha particles \citep[e.g.][]{Phan.GRL.2013}, we can infer the Alfv\'en speed at the source $\frac{1}{2} m_\alpha v_A^2 \sim$ 85 keV, which gives $v_A \sim$ 2020 km/s, not inconsistent with expections at low altitudes \citep[e.g.][]{1999SSRv...87...25A,Warmuth.AstronomyAstrophysics.2005}.  \cite{2009ApJ...700L..16D} describe simulations of a ion pickup process in reconnection outflow regions that effectively energizes alpha particles to  $\sim \frac{1}{2} m_\alpha v_A^2$.

The observation that the lower-$|B|$ funnels are replete with switchbacks is consistent with the theory that switchbacks grow as a consequence of Alfv\'enic fluctuations reaching large amplitudes through solar-wind expansion \citep{2020ApJ...891L...2S,2021arXiv210408321M}. Specifically, funnel regions have presumably undergone more super-radial expansion than neighbouring regions, so even minor variations in relative fluctuation amplitudes at the source surface could lead to large differences at the location of PSP \citep{Hollweg1974}. Fluctuations that grow to $\delta B/B\gtrsim 1$ then form switchbacks or potentially fluxropes associated with turbulent reconnection \citep[e.g.][]{2021A&A...650A...2D}.  The relation of this to other solar-wind properties (wind speed, proton temperature, and alpha fraction) remains an open problem: the funnels described here lead to solar wind streams dubbed Alfv\'enic slow streams \citep{Stansby.MNRAS.2019, 2021JGRA..12628996D}, rather than the fast wind typically associated with polar coronal hole outflows. The reason is that the speed of the solar wind is controlled not by the local expansion discussed here occurring within 20 Mm from the photosphere, but the overall coronal hole expansion. PSP, always skirting the current sheet in the ecliptic plane as the Sun's activity picks up out of the latest minimum, has been mostly observing wind coming from rapidly expanding open field lines. For this type of wind, the heating and acceleration is similar to fast solar wind \citep{Chandran2021}, including plasma properties, but the global coronal geometry determines the slower acceleration profile \citep{Wang1990, 2019ApJ...873...25P}.

The stream structure and fluctuations shown here must evolve to become the familiar solar wind at $1~au$ and beyond. \cite{Horbury2021} have used measurements by Solar Orbiter of the same solar wind stream at $200~R_S$ as presented here: while the alpha particle modulation was preserved, the longitudinal speed variations were smoothed out by that distance, leading to density variations on the same longitudinal scale. The clear variation in fluctuation power due to switchback modulation that was present at $25~R_{S}$ was replaced by a large scale magnetic field variation on the same scale, suggesting that the photospheric structure affecting the wind at PSP still retains a signature in the wind much farther from the Sun.

Coronal magnetic activity is naturally multi-scale.  Large-scale activity (e.g., flares and CMEs) correlate well with the sunspot solar cycle, while smaller-scale activity (e.g., bright-points, plumes, and jets) is more ubiquitous regardless of the solar cycle phase.  Space and ground-based observations in the 1970s provided the first evidence for the highly dynamic nature of the coronal base \citep{1973SoPh...31..449D,1975ApJ...197L.133B,1976ApJ...203..528W,1980ApOpt..19.3994B}. Yohkoh/SXT observations showed most energetic coronal jets \citep[e.g.,][]{1992PASJ...44L.173S,1992PASJ...44L.161S,1996PASJ...48..123S,1998SoPh..178..379S,2001ApJ...550.1051S}.  These discoveries led to speculations on the role these transients play in the heating and acceleration of the solar wind plasma \citep{1983ApJ...272..329B}.

Our {\em in situ} measurements imply a connection between impulsive Alfv\'enic activity within discrete wind sources and solar wind heating and acceleration mechanisms.  
While our measurements appear to be consistent with interchange reconnection, either through direct heating from the reconnection site itself or through enhanced Poynting fluxes (e.g., waves) that dissipate at higher altitudes, there are other viable mechanisms to consider. 
\cite{1999SSRv...87...25A} describe a wave-heating process occurring within magnetic funnels:  high frequency Alfv\'en waves propagating outward enter resonance with ions and heat perpendicular to the local magnetic field and the resulting mirror force produces accelerated flow \citep{1998A&A...335..303C}; resonant ion cyclotron heating is observed at 1 au \citep{2013PhRvL.110i1102K}.   \cite{MartinezSykora2017} suggest that ambipolar diffusion transports kinked field lines near network field concentrations into the chromosphere and the stored magnetic tension is released impulsively to form type II spicules and Alfv\'enic jets.

Finally, it is interesting to speculate about a connection to coronal plumes - persistant filamentary structures that disappear into the general solar wind outflow several tens of solar radii above the Sun's surface \citep{2011A&ARv..19...35W,Poletto2015, 2016SSRv..201....1R, 10.1023/a:1004955223306,2003ApJ...588..566T,10.1007/s11207-008-9171-2,2014ApJ...784..166Y}. Plumes are also known to emerge from network field concentrations with enhanced heating at the base and are characterized by higher density \citep{10.1086/187617} and lower electron temperatures \citep{10.1023/a:1004955223306} than the surrounding corona. \cite{2008ApJ...682L.137R} show that coronal plumes display myriads of small, short-lived jets at their footpoints \citep{2014ApJ...787..118R}, most probably resulting from magnetic reconnection due to small emerging bipolar features \citep[see e.g.,][]{2018ApJ...868L..27P,2019ApJ...887L...8P}. 
Plumes exhibit oscillations and sub-structures \citep{2021ApJ...907....1U} and should expand in pressure balance into the solar wind \citep{2011ApJ...736...32V}, but their potential relationship to the structures observed by PSP and reported here remains to be explored.

\acknowledgments
Parker Solar Probe was designed, built, and is now operated by the Johns Hopkins Applied Physics Laboratory as part of NASA's Living with a Star (LWS) program (contract NNN06AA01C). Support from the LWS management and technical team has played a critical role in the success of the Parker Solar Probe mission.
We would like to acknowledge the efforts of the FIELDS, SWEAP, and ISOIS instrument and science operations teams and the PSP spacecraft engineering 
and operations team at the Johns Hopkins Applied Physics Laboratory. O. Panasenco and M. Velli were partially supported by the HERMES DRIVE NASA Science Center grant No. 80NSSC20K0604.  B. Chandran was supported by NASA grants NNX17AI18G and 80NSSC19K0829.  J.~Squire was supported by the Royal Society Te Ap\=arangi NZ through Rutherford Discovery Fellowship RDF-U00180. T. Horbury and L. Woodham were supported by UK STFC grant ST/S000364/1, T. Woolley by ST/N504336/1 and R. Laker by an Imperial College President's scholarship.

\bibliography{funnels}{}
\bibliographystyle{aasjournal}

\end{document}